\documentclass[submission,copyright,creativecommons]{eptcs}
\providecommand{\event}{ICLP/LPNMR-DC 2024} 

\usepackage{iftex}

\ifpdf
  \usepackage{underscore}         
  \usepackage[T1]{fontenc}        
\else
  \usepackage{breakurl}           
\fi

\usepackage{booktabs}
\usepackage{dsfont}
\usepackage{wrapfig} 
\usepackage{amsmath}
\usepackage{graphicx}
\usepackage{multirow}
\usepackage{multicol}
\usepackage{xspace}
\usepackage{xcolor}
\usepackage{comment}    
\usepackage{listings}
\usepackage[most]{tcolorbox}

\definecolor{noteColor1} {rgb}{  1,  0,  0} %
\definecolor{noteColor2} {rgb}{  1,0.5,  0} %
\definecolor{noteColor13}{rgb}{0.9,0.2,  0} 
\definecolor{noteColor3} {rgb}{0.8,0.8,  0}
\definecolor{noteColor4} {rgb}{  1,  0,0.6}
\definecolor{noteColor5} {rgb}{  1,  0,  1} %
\definecolor{noteColor6} {rgb}{  0,0.8,  0} %
\definecolor{noteColor7} {rgb}{  0,0.8,0.6} %
\definecolor{noteColor8} {rgb}{  0,0.8,0.8} %
\definecolor{noteColor9} {rgb}{0.5,0.8,  0} %
\definecolor{noteColor10}{rgb}{  0,  0,  1} %
\definecolor{noteColor11}{rgb}{0.5,  0,  1} %
\definecolor{noteColor12}{rgb}{  0,0.5,  1} %

\definecolor{maroon}{rgb}{0.5,0,0}
\definecolor{darkgreen}{rgb}{0,0.5,0}
\definecolor{mildgreen}{rgb}{0,0.8,0}
\definecolor{lightblue}{rgb}{0.3,0.3,1}
\definecolor{lightgrey}{rgb}{0.97,0.97,0.97}
\definecolor{grey}{rgb}{0.65,0.65,0.65}

\iftrue  
  \newcommand{\TODO}   [1]{\textcolor{noteColor1}{\textbf{\small[[TODO: #1]]}}\\}
  \newcommand{\TODOL}  [1]{\textcolor{noteColor1}{\textbf{\small[#1]}}\\}
  \newcommand{\todo}   [1]{\textcolor{noteColor1}{\textbf{\small[[TODO: #1]]}}} 
  \newcommand{\todos}   [1]{\textcolor{noteColor1}{\textbf{[[TODO: #1]]}}}      
  \newcommand{\todol}  [1]{\textcolor{noteColor1}{\textbf{\small[#1]}}}	      

  \newcommand{\hide}   [1]{\textcolor{grey}       {#1}}
\else
  \newcommand{\TODO}   [1]{}
  \newcommand{\TODOL}   [1]{}
  \newcommand{\todo}  [1]{}
  \newcommand{\todos}  [1]{}
  \newcommand{\todol} [1]{}
  
  \newcommand{\hide}   [1]{}

\fi

\newcommand{\scasp}[0]{s(CASP)}
\newcommand{\clingo}[0]{\textsc{Clingo}}
\newcommand{\clingolp}[0]{\textsc{Clingo[LP]}}
\newcommand{\hilite}[0]{\textsc{HiLiTE}}

\lstnewenvironment{lstlistingwolinenum}{
    \lstset{
        basicstyle=\ttfamily\footnotesize,
        frame=lrtb,
        breaklines=true
    }
}{}
\lstnewenvironment{lstlistingwlinenum}{
    \lstset{
        basicstyle=\ttfamily\footnotesize,
        frame=lrtb,
        breaklines=true,
        numbers=left,
        numbersep=0.5em,
        xleftmargin=1.5em,
        framexleftmargin=1.5em,
        stepnumber=1
    }
}{}

\usepackage{adjustbox}          
\usepackage{relsize}            
\definecolor{PrologPredicate}{RGB}{0,0,200}
\definecolor{PrologVar}      {RGB}{145,032,039}
\definecolor{PrologComment}  {RGB}{0,170,0}
\definecolor{PrologOther}    {rgb}{0.2,0.2,0.2}
\definecolor{PrologString}   {RGB}{070,120,200}
\usepackage{textcomp}

\lstdefinestyle{MyInline}
{
  basicstyle = \relsize{-0.5}\ttfamily\color{PrologPredicate},
  escapechar = @,
  escapeinside = {-<}{>-},
  breaklines = true,
  breakatwhitespace=true,
  upquote = true,
  literate =
  {?-}{{?-\,}}3
  {:-}{{:-\,}}3
  {.=.}{{\,\#=\,}}3
  {.<.}{{\,\#<\,}}3
  {.>.}{{\,\#>\,}}3
  {.=<.}{{\,\#=<\,}}4
  {.>=.}{{\,\#>=\,}}4
}
\lstdefinestyle{MySCASP}
{
    numbersep=1em,    
    xleftmargin=0.45cm,  
    numberstyle=\tiny,
    numbers=left,
    stepnumber=1,
  mathescape = true,
  escapechar = @,
  escapeinside = {-<}{>-},
  keywords = {},
  upquote = true,
  basicstyle = \ttfamily\relsize{-.5}\color{PrologPredicate},
  basewidth = 0.47em,
  moredelim = {*[s][\color{PrologVar}]{(}{)}},
  moredelim = {*[s][\color{PrologString}]{'}{'}},
  moredelim = {*[s][\color{PrologOther}]{:-}{.}},
  commentstyle = \mdseries\color{PrologComment},
  escapebegin=\color{PrologVar},
  morecomment=[l]\%,
  literate     =
  {|}{{|}}2
  {.=.}{{\,\#=\,}}3
  {.<.}{{\,\#<\,}}3
  {.>.}{{\,\#>\,}}3
  {.=<.}{{\,\#=<\,}}4
  {.>=.}{{\,\#>=\,}}4
}

\newtcolorbox{codeBox}{%
    colframe=black!75,
    enhanced,
    use color stack,
    breakable,
    left=2pt, right=0pt, top=-4pt, bottom=-4pt, boxsep=0pt, boxrule=0.25pt, arc=2pt,
    bottomsep at break=6pt,
    topsep at break=6pt,
    coltitle=white, 
    detach title,
}
\newtcolorbox{enumBox}{%
    colframe=black!75,
    enhanced,
    use color stack,
    breakable,
    left=2pt, right=2pt, top=0pt, bottom=0pt, boxsep=2pt, boxrule=0.25pt, arc=2pt,
    bottomsep at break=0pt,
    topsep at break=0pt,
    coltitle=white, 
    detach title,
}

\title{Early Validation of High-level Requirements\\ on Cyber-Physical Systems}
\author{Ond\v{r}ej Va\v{s}\'{i}\v{c}ek
\institute{Faculty of Information Technology\\
Brno University of Technology\\ 
Brno, Czechia}
\email{ivasicek@fit.vut.cz}
}
\def\titlerunning{Early Validation of High-level Requirements on Cyber-Physical Systems}
\def\authorrunning{Ondřej Vašíček}

\usepackage{pdfpages}

\begin{document}
\maketitle


\begin{abstract}
The overarching, broad topic of my research are advancements in the area of \emph{safety-critical}, \emph{cyber-physical} systems (CPS) development with emphasis on validation and verification.
The particular focus of my research is the early validation of high-level requirements on CPS. 
My current approach for tackling this problem is transforming the requirements into Event Calculus and subsequently reasoning about them using ASP solvers such as the grounding-free \scasp{}.
Below, I discuss my research, its current state, and the open issues that are still left to tackle.
The first results of my work will be presented in a~paper that was accepted for ICLP'24, which is my first paper in this area.

\end{abstract}

\section{Introduction and Motivation}\label{intro}

Early validation of specifications describing requirements placed on cyber-physical systems (CPSs) under development is essential to avoid costly errors in later stages of the development, especially when the systems undergo certification.
According to the study~\cite{avsi_savi} conducted within the AVSI SAVI project around~70~\% of errors in large systems
are introduced during the specification of system requirements, yet over~50~\% of those errors are only discovered
during the integration testing phase much later in the development process. 
Unfortunately, the cost to fix an error in the integration testing phase is around 16-times higher than in the initial requirements
specification phase. 
%

To our best knowledge there are currently no ready-to-use solutions for automated, early validation of
truly high-level requirements.
Requirements specifications are still most often in textual form which traditionally suffers
from ambiguity and is not easily machine understandable. 
The most common way of validating such requirements is expert review, which can lead to errors being overlooked
due to the human factor, due to the size and complexity of requirements specifications, and due to implicit assumptions
not being included explicitly in the specification.
There is a~need for better ways to express clear and unambiguous requirements, to relate them to system model elements,
and to be able to automatically transform both the requirements and the models into suitable formalisms which could
then be used by methods for validation and verification.

A~crucial need, when trying to transform a~requirements specification into a~suitable formalism, is that of a~small semantic gap between the requirements and the formalism used to model them for the purposes of validation.
A~larger semantic gap makes it more difficult to transform the requirements into a~model, and, most importantly, any validation on such a~model drifts away from validating the requirements themselves and closer to validating that particular model---influenced by design and implementation decisions.
As described in~\cite{mueller_book-fixed}, Event Calculus (EC) is a~formalism suitable for commonsense reasoning.
The semantic gap between a~requirements specification and its EC encoding is near-zero because its semantics follows how a~human would think of the requirements.
Using Answer Set Programming (ASP)~\cite{asp} and the \scasp{}~\cite{scasp} system for goal-directed reasoning in EC,
the work \cite{gupta-train} has demonstrated the versatility of EC for modelling and reasoning about CPSs while providing explainable results.
My latest research~\cite{iclp-ec-scasp-pcapump} builds on~\cite{gupta-train} in order to further enhance its capabilities and potential applications.

\section{Background}
\paragraph{The Event Calculus (EC)~\cite{mueller_book-fixed}} EC is a~formalism for reasoning about events and
change, of which there are several
axiomatizations.  There are three basic concepts in EC:
\emph{events}, \emph{fluents}, and \emph{time points}:
(i) an event is an action or incident that may occur in the world, e.g., the dropping of a~glass by a~person is an event,
(ii) a~fluent is a~time-varying property of
the world, such as the altitude of a~glass,
(iii) a~time point is an instant of time.
Events may happen at a~time point; fluents have a~truth value at any
time point
, and these truth values are subject to
change upon an occurrence of an event.  In addition, fluents may have
quantities associated with them as parameters, 
which change discretely via events or continuously over time via trajectories.
We chose EC as a~formalism suitable for representing requirements specifications due the the low semantic gap between EC and the requirements.

\paragraph{The \scasp{} System~\cite{scasp}} \scasp{} extends the
expressiveness of ASP~\cite{asp} systems, based on the stable
model semantics~\cite{gelfond88:stable_models-fixed}, by including
predicates, constraints among non-ground variables, uninterpreted
functions, and, most importantly, a~top-down, query-driven execution
strategy.
These features make it possible to return answers with non-ground
variables (possibly including constraints among them) and to compute
partial models by returning only the fragment of a~stable model that
is necessary to support the answer to a~given query.
Answers to all queries can also include the full proof tree, making them fully explainable.
Like other ASP implementations and unlike Prolog, s(CASP) handles
non-stratified negation and returns the corresponding (partial) stable
models.
Further, \scasp{} implements abductive reasoning via even loops, where it automatically searches for suitable values of the predicates in the corresponding even loop in order to satisfy the main query.
We chose \scasp{} as the solver for reasoning about EC models especially due to its grounding-free nature, which allows us to reason in continuous time and about continuous change of fluents.

\section{Related Work}

As already mentioned, we found EC suitable for reasoning about requirements specifications due to its low semantic gap against them.
In comparison, the semantics of automata-based approaches, which are often used in the literature to model CPSs, such as timed automata~\cite{timed-automata-uppaal} or hybrid automata~\cite{hybrid-automata}, require one to ``design" explicit states and transitions, and may lead to decomposition of the system into sub-systems each with their own automaton.
Current industrial model-based engineering approaches, such as those based, e.g., on Matlab Simulink models and tools like \hilite{}~\cite{hilite}, are only suitable for validation of low-level requirements.
This is due to the low-level nature of the models they use, especially when automated generation of code from the models is required.
Much research has been done on using temporal logics (e.g., LTL, CTL, CTL*~\cite{hand-book-modelchecking}) and real time temporal logics (e.g., MTL~\cite{temporal-logic-mtl}) to represent system properties.
However, we have not considered temporal logics as a~target for transforming high-level system specifications since the semantics of the temporal logics that we are aware of are further away from natural language which makes the transformation more difficult to perform and to understand in comparison with EC.

Apart from the EC-based approach introduced in~\cite{gupta-train}, which my work builds upon, 
there are other ones which aim to target automated validation of high-level requirements put on CPSs.
The work~\cite{ge-assert} is based on ontologies and uses theorem proving, which traditionally requires significant manual work.
The work~\cite{cea-glossaries-and-process-algebras} is based on transforming CPS specifications from templated-English
into process algebras extended with real-time aspects, however, no continuous variables (apart from time) are dealt with and no experimental results are presented, which makes it difficult to judge the scalability of this approach.

Finally, there are other ASP solvers than the \scasp{} system that I am currently using, such as the grounding-based \clingo{}~\cite{clingo-paper}.
However, in my experiments so far, \clingo{} has, unfortunately, proven to be unsuitable for reasoning about fluents with large or continuous value domains due to the explosion in the grounding and a~need to discretize the time.
This causes the solver to quickly run out of memory even on models with very limited continuous value domains, while the discretized time steps introduce issues when trying to step on exact values during periods of continuous change.
An advantage of \clingo{} is that it does not suffer from non-termination issues, which make things much more complicated in \scasp{}, therefore, I plan to further investigate the possibility of using it in my research once advancements are made in reducing the grounding explosion.

\section{Research Goals and Current State}

The overarching objective of my research is to improve the development process of safety-critical systems
with a~focus on validation and verification.
The main concrete goal is to propose analyses for high-level requirements in order to detect more errors as early
as possible in the development process.
This goal consists of three sub-goals.
In my research thus far, I have already partially completed the first two sub-goals and I am recently focusing on the third one.
\begin{enumerate}
    \item
    A~suitable formalism needs to be selected which will be able to adequately
    represent requirements specifications for the purpose of performing analyses on the requirements.
    In my research, I have selected Event Calculus as a~suitable formalism based on already published results~\cite{gupta-train,midas-eventcalc-asp} and on my own experiments.
    Experimenting with the practical capabilities of EC is an ongoing effort entailed with sub-goal~3.
    
    \item
    Then, a~way of transforming the requirements specifications into the selected formalism needs to be defined.
    For EC, limited manual transformations have already been shown in~\cite{gupta-train,midas-eventcalc-asp}.
    In my research, I have since used further manual transformations for the time being.
    Once I sufficiently explore and advance the modelling and reasoning capabilities of EC, my focus will shift towards proposing a~general and at least semi-automated way to transform requirements into EC.
   
    \item
    Further research should focus on efficient analysis methods for validating the formalized requirements specifications.
    The aim is to make the reasoning more scalable, to make it capable of supporting more constructs
    from the requirements specifications, and to use it for new kinds of analyses.
    Since my selected formalism is currently EC, my recent research is focusing on the capabilities of abductive commonsense reasoning using ASP solvers~\cite{scasp,clingo-paper}, which have already been shown to be promising by other researchers in the past~\cite{gupta-train, midas-eventcalc-asp}.
\end{enumerate}

\noindent
The next section discusses the first published result of my research.
The work uses EC (from sub-goal~1) and a~manual transformation (from sub-goal~2) in order to explore and improve the reasoning capabilities of ASP solvers, especially \scasp{}, for the purposes of requirements validation (towards sub-goal~3).

\section{Current Results}

In my research so far, I have studied EC and experimented with \clingo{} and \scasp{}.
In my latest work, I chose to use \scasp{} because it does not suffer from the grounding explosion.
In order to both assess and demonstrate the practical capabilities (and current limitations) of the EC+\scasp{} approach and to guide my work on improving its capabilities, I have applied it on the specification of a~real safety-critical system.
I will be presenting the results of this endeavour in a~paper which was accepted at ICLP'24~\cite{iclp-ec-scasp-pcapump} as my first paper in this area of research. 
The paper is briefly summarized below.
\\

We develop a~model of the core operation of the PCA pump~\cite{pcapump-paper}---a~real safety-critical device that automatically delivers pain relief drugs into the blood stream of a~patient.
The model operates in a~way similar to an early prototype of the system and, thus, can be used to reason about its behaviour.
However, due to the nature of EC, the behaviour of the model is very close to the behaviour described by the requirements themselves.
%
%
The requirements specification and all the source codes of its s(CASP) representation can be found at \url{https://github.com/ovasicek/pca-pump-ec-artifacts/}.
The below is a~brief, illustrative overview of s(CASP) code for the delivery of a~patient bolus (one of the features of the pump), which is an extra dose of drug delivered upon the patient's request.
We define events that start and end the delivery of the bolus which is represented by a~state fluent (lines 1-4).
The total amount of drug delivered to the patient and how the amount increases during the bolus is represented by a~continuous fluent and a~trajectory (lines 6-10).
And finally, the bolus stops automatically once the right amount of drug (so called VTBI) is delivered which is represented by an event triggered based on the drug delivered during the ongoing bolus (lines 12-15).
A new continuous fluent and trajectory are used to represent the volume of drug delivered by a~bolus counting from zero instead of computing the difference of total drug delivered at the start and at the end of a bolus (lines 17-20).
\vspace{-0.5em}
\begin{codeBox}
\begin{lstlisting}[style=MySCASP]
fluent(patient_bolus_delivery_enabled).
event(patient_bolus_delivery_started).   event(patient_bolus_delivery_stopped).
initiates(patient_bolus_delivery_started, patient_bolus_delivery_enabled, T).
terminates(patient_bolus_delivery_stopped, patient_bolus_delivery_enabled, T).

fluent(total_drug_delivered(X)).
trajectory(patient_bolus_delivery_enabled, T1, total_drug_delivered(Total), T2) :-
  basal_and_patient_bolus_flow_rate(FlowRate),
  holdsAt(total_drug_delivered(StartTotal), T1),
  Total .=. StartTotal + ((T2 - T1) * FlowRate).
  
event(patient_bolus_completed).
happens(patient_bolus_completed, T2) :-   initiallyP(vtbi(VTBI)),
  holdsAt(patient_bolus_drug_delivered(VTBI), T2).
happens(patient_bolus_delivery_stopped, T) :- happens(patient_bolus_completed, T).

fluent(patient_bolus_drug_delivered(X)).
trajectory(patient_bolus_delivery_enabled,T1, patient_bolus_drug_delivered(X),T2):-
  patient_bolus_only_flow_rate(FlowRate),
  X .=. (T2 - T1) * FlowRate.
\end{lstlisting}
\end{codeBox}

%
The first validation method that we propose is a~way to check the consistency between the behaviour defined by the requirements specification and the use cases (UC) and exception cases (ExC) based on which the requirements were created (or, in general, checking consistency of the behaviour against any scenarios defined at a~different level of the specification).
This is done by transforming the UC/ExC into an EC narrative and forming a~query based on the UC/ExC and its post-conditions.
If running the query on the narrative using \scasp{} fails, then we have found an inconsistency.
Using this technique we were able to identify a~number of such inconsistencies in the PCA pump specification.


The second validation method that we propose is a~way to check whether the requirements specification satisfies general properties, such as that the system should not allow an overdose of the patient or that the system should respond to an event within a~given time limit.
This is done by representing a~general property as an \scasp{} query and checking that query on suitable narratives.
In this way, we were able to detect that the patient can be overdosed by the PCA pump in certain specific narratives, which is a~safety property violation caused by a~missing requirement.
We were further able to leverage the abductive reasoning capabilities of \scasp{} in order to generalize the narrative on which the property is being checked.
In our case of checking the possibility of an overdose, we were able to abduce the parameters of an overdose (what volume of drug is allowed over what time period) and, subsequently, detect the possibility of an overdose in a~``sunny day'' narrative (in which the overdose does not directly occur otherwise).

We further present a~number of challenges encountered during the translation of the requirements to EC encoded in s(CASP) and during the subsequent evaluation, based on deductive as well as abductive reasoning, which was often too costly or non-terminating.
We have applied and, in multiple cases, also newly developed various techniques that helped us resolve many of these challenges.
These include extensions of the axiomatization of the EC and special ways of translating certain parts of the specifications in order to avoid non-termination.
Further, we present an original approach to abductive reasoning in \scasp{} with incrementally refined abduced values in order to assure consistency of the abduced values whenever abduction on the same value is used multiple times in the reasoning tree. 
Next, we proposed a~mechanism for caching predicate evaluations (failure-tabling and tabling of ground sub-goal success) that was added into \scasp{} as a~prototype leading to a~significant increase in performance.
We also describe a~way of separating the reasoning about the trigger and the effect of certain complexity-inducing triggered events into multiple reasoning runs where each run produces new facts to be used in the subsequent ones, which reduces their performance impact.
\\

Our work demonstrated that EC can be used to model the requirements specification of a~non-trivial, real-life cyber-physical system in \scasp{} 
and the reasoning involved can lead to discovering issues in the requirements while producing valuable evidence towards their validation.
Indeed, the work resulted in the discovery of a~number of issues in the PCA pump specification, which we have discussed and confirmed with the authors of the specification.

\section{Open Issues}

There are still many open issues to be tackled in future work.
A range of them consists of implementation tasks needed to improve the \scasp{} system, such as properly integrating and efficiently implementing our abductive reasoning semantics directly within \scasp{} and further improvements to our prototype caching.
Such issues are less interesting from the research perspective due to their engineering nature.

\paragraph{\scasp{} Non-termination} The main issue that is currently limiting \scasp{} and our use of its reasoning capabilities is non-termination, which is often related to Zeno behaviours caused by reasoning in continuous time.
A common non-termination case is the ``toggle" scenario where a~system toggles between two fluents affected by respective toggle events. 
In such cases, the s(CASP) reasoner without guidance, currently fails to resolve the time intervals when each of the fluents may or may not hold even with a~single toggle event.
Another source of non-termination is abduction of timepoints of event occurrences, because \scasp{} will attempt to abduce an infinite number of events due to reasoning in continuous time.
In certain cases, we are able to avoid non-termination by manually modifying the way \scasp{} handles negated predicates, however, we lack a~general solution to the problem.
We believe that this issue could be solved in some cases via loop detection techniques.
Another direction of research, which could diminish this issue, is the creation of a~meta-reasoner in \scasp{} specialized to EC, which would be more efficient and better at avoiding non-termination by leveraging the knowledge of the semantics of EC axioms and predicates. \scasp{} is currently in no way specialized for EC.

\paragraph{Performance scaling} Another issue is performance scaling of \scasp{} reasoning.
Even with our prototype cache, which greatly reduces reasoning runtime, we still observe a~steep exponential increase in solving complexity when introducing more events into a~narrative (specifically input events).
We believe that the scaling can be greatly improved via a~more efficient approach to reasoning.
One source of inefficiency is the the need to prove the entire history of the timeline from time zero up to the current time whenever we reason about any aspects of EC at a~given timepoint.
We believe that many predicates are being re-proven multiple times (instead of just once) throughout the process of answering a~query due to the need to re-prove the entire history in order to prove anything.
This was partially addressed by our prototype cache, however, the scaling currently still remains exponential.
A promising approach, which has already been used in our ICLP paper in a~very limited form, is incremental solving where we execute restricted reasoning multiple times and transfer facts between runs.
My latest experiments with a~generalization of this incremental approach, which only reasons about smaller intervals of the narrative timeline at a~time, have shown a~potential to significantly improve the performance scaling on fixed narratives (i.e., narratives with at most one solution).

When considering \clingo{} as a~solver, we observe a~very steep (most likely exponential) increase in both solving time and memory consumption when increasing the value domain of variables (such as the time scale, drug volume, etc.) due to an explosion in grounding size.
My efforts are currently mainly focused on \scasp{} and I have a~deeper understanding of its capabilities due to my collaboration with J.~Arias and G.~Gopal, who are the authors of \scasp{}. 
In comparison, I only have limited experience with and knowledge of \clingo{}.
I would be very interested in any ways to reduce the grounding explosion when reasoning about systems which use continuous time and variables.
A possible solution might prove to be theory reasoning extensions of \clingo{}~\cite{clingo-theory}, such as \clingolp{}~\cite{clingolp}.

\paragraph{Generality and Automation}
Further open issues are related to making our approach more general and practically usable.
These belong more into the Software Engineering Engineering research domain, rather than in the domain of Logic Programming.
Currently, our transformation of requirements to EC is entirely manual, although we try to keep it as general as possible.
In order for our approach to be practically usable in the industry, we need to propose at least a~semi-automated transformation for general requirements of a~wide enough class of systems.
The main obstacle in the way of automation is the format of requirements specifications as they are largely written in unconstrained natural language.
Natural language is inherently ambiguous and hard to reliably process by machines. 
We believe that introducing a~more structured language, such as MIDAS by~\cite{midas}, for facilitating formulation of requirements should provide enough structure and context to the requirements in order to enable a~more general and at least semi-automated transformation of the requirements into EC.
Another option to tackle this issue might be the use of LLMs due to their recent success in processing natural language, however, the issue of hallucinations and lack of explainability makes their use difficult in the domain of safety-critical systems which have to go through certification.

\bibliographystyle{eptcs}
\bibliography{general}
\end{document}